\documentclass[%
 reprint,
superscriptaddress,
groupedaddress,
 amsmath,amssymb,
aps,
pra
]{revtex4-2}

\usepackage{graphicx}
\usepackage{dcolumn}
\usepackage{bm}
\usepackage{xcolor}
\usepackage[mathlines]{lineno}

\begin{document}


\title{Model for Predicting Adsorption Isotherms and the Kinetics of Adsorption via Steepest-Entropy-Ascent Quantum Thermodynamics    }
\author{Adriana Saldana-Robles}
\altaffiliation[Also at the ]{Department of Agricultural Engineering, University of Guanajuato.}
  \email{adrianarobl@vt.edu}
\author{Cesar Damian}%
\altaffiliation[Also at the ]{Department of Mechanical Engineering, University of Guanajuato.}
 \email{cesarasce@vt.edu}
  \author{William T. Reynolds Jr.}%
 \email{reynolds@vt.edu}
\author{Michael R. von Spakovsky}%
 \email{vonspako@vt.edu}
 \affiliation{
Center for Energy System Research, Mechanical Engineering Department, Virginia Tech, Blacksburg, VA 24061
}%



\date{\today}

\begin{abstract}
This work outlines the foundations for being able to do a first-principle study of the adsorption process using the steepest-entropy-ascent quantum thermodynamic (SEAQT) framework, a framework able to predict the unique non-equilibrium path taken by a system from some initial state to stable equilibrium. To account for the process of multi-component adsorption, the SEAQT framework incorporates the particle number operator for each absorbed species directly into its equation of motion. The theoretical models developed are validated via  some initial comparisons with experimental data found in the literature, demonstrating good agreement. The findings reveal that this framework can be an effective tool for describing the adsorption process out of equilibrium. It is able to do so without $a \; priori$ knowledge of the specific adsorption mechanism(s) involved. It also aligns well with the anticipated predictions of equilibrium models. In addition, within this framework, all intensive thermodynamic properties are characterized by out-of-equilibrium fluctuations, highlighting the significance of non-equilibrium thermodynamics in predicting measurable physical quantities.

\end{abstract}

\maketitle
\textit{}

\section{Introduction}
The quality of water is fundamental for both human and ecological health. However, water pollution has been increasing from numerous chemical compounds including dyes, organic compounds, and metals originating from both industrial and natural sources. As a consequence, many water treatment methods have been implemented to remove contaminants and meet rigorous safety standards~\cite{anastopoulosAreThermodynamicParameters2016,ghosalDevelopmentGeneralizedAdsorption2017}. Adsorption stands out as one of the most viable methods for removing contaminants from water due to its cost-effectiveness. As a result, extensive research has been conducted in recent decades to develop improved adsorbent materials~\cite{degisiCharacteristicsAdsorptionCapacities2016, jeiraniAdsorptionEmergingPollutants2017, carvalhoAdsorbateadsorbentPotentialEnergy2022}.

Beyond material development, there is a growing interest in understanding the mechanisms and thermodynamics of adsorption~\cite{carvalhoAdsorbateadsorbentPotentialEnergy2022}. In particular, adsorption isotherms are crucial for elucidating the pathways and mechanisms of adsorption on the adsorbent surfaces~\cite{ayaweiModellingInterpretationAdsorption2017, cherkasovLiquidphaseAdsorptionCommon2020}. However, the information obtained is only at stable equilibrium and requires trial-and-error regression analysis to determine the most suitable model for describing the experimental data.

Several equilibrium adsorption models have been proposed in the past to elucidate the pathways and thermodynamics of adsorption~\cite{anastopoulosAreThermodynamicParameters2016}. These models can be categorized as a) one-parameter models such as Henry's isotherm~\cite{faustAdsorptionProcessesWater2013}; b) two-parameter models, which include the Hill-Deboer~\cite{ayaweiModellingInterpretationAdsorption2017}, Fowler-Guggenheim~\cite{sampranpiboonEquilibriumIsothermModels2014}, Langmuir isotherm~\cite{elmorsiEquilibriumIsothermsKinetic2011}, Freundlich isotherm~\cite{ayaweiSynthesisCharacterizationApplication2015}, Dubinin-Radushkevich isotherm~\cite{ayaweiModellingInterpretationAdsorption2017}, Temkin isotherm~\cite{ringotVitroBiosorptionOchratoxin2007} and other models; c) three-parameter models such as the Redlich-Peterson isotherm~\cite{davoudinejadModelingAdsorptionIsotherm2013}, Sips isotherm~\cite{jeppuModifiedLangmuirFreundlichIsotherm2012}, Toth isotherm~\cite{behbahaniNewStudyAsphaltene2014}, Kahn isotherm~\cite{ayaweiModellingInterpretationAdsorption2017}, and Langmuir-Freundlich isotherm~\cite{ayaweiModellingInterpretationAdsorption2017} models; and d) four-parameter models such as the Fritz-Schlunder isotherm~\cite{yanevaLinearNonlinearRegression2013}, Baudu isotherm~\cite{mckayOptimumIsothermsDyes2014}, and Weber-Van Vliet isotherm~\cite{ayaweiModellingInterpretationAdsorption2017} models. However, many of these models are empirical and provide limited information about the adsorption process~\cite{ghosalDevelopmentGeneralizedAdsorption2017, carvalhoAdsorbateadsorbentPotentialEnergy2022} making it challenging to select the appropriate model. 

Because of these challenges, the adsorption process has also been modeled using neural networks and adsorbate-adsorbent potential energy functions based upon second virial coefficient data. This approach has been used to evaluate energy interactions and the geometric optimization between the adsorbate and adsorbent~\cite{carvalhoAdsorbateadsorbentPotentialEnergy2022}. Like traditional isotherm models, such an optimization focuses solely on adsorption at equilibrium and it is unable to provide information about the non-equilibrium process. This lacuna occurs because a description of adsorption kinetics requires $a \; priori$ knowledge of the reaction mechanisms.

Various models have been developed to understand the dynamics of non-equilibrium adsorption; these seek to determine the time required for the adsorption process. Models such as: a) the pseudo-first-order rate equation, b) the pseudo-second-order rate equation, c) Elovich’s equation, d) the liquid film diffusion model, e) the intraparticle diffusion model, and f) the double-exponential model (DEM) are widely employed for fitting kinetic data. The main challenge with models a), b), and c) is that adsorption reaction rate models must be chosen depending upon assumed adsorption mechanisms. However, because they do not represent the course of the actual adsorption process, they cannot provide useful information about the adsorption mechanism involved. In contrast, models d), e), and f) are able to represent the course of the adsorption process in a more reasonable manner~\cite{qiuCriticalReviewAdsorption2009}.

Of course, all of these equilibrium and non-equilibrium approaches necessarily describe the adsorption process under limiting conditions (e.g., dilute solutions, a single adsorbing species, uniform surface properties, monolayer adsorption, etc.), and they lack the ability to account for the competing adsorption mechanisms present in complex systems. A truly predictive model should be able to describe the adsorption kinetics from the underlying energetic and entropic principles without overly simplistic assumptions. In this regard, the present manuscript proposes using the steepest-entropy-ascent quantum thermodynamic (SEAQT) framework to describe the adsorption process both at equilibrium and non-equilibrium. Without assuming any particular reaction mechanism, it aligns well with the predictions of known isotherm models as well as those of kinetic models but provides results that go well beyond these models, particularly in being able to treat complex multi-component systems.

The SEAQT framework is built upon fundamental principles of thermodynamics and quantum mechanics, and it is able to predict the unique non-equilibrium path taken by a thermodynamic adsorbing system. It does so with the principle of steepest entropy ascent (SEA)~\cite{berettaSteepestentropyascentNonequilibriumQuantum2017, kusabaModelingNonEquilibriumProcess2017, mcdonaldPredictingNonequilibriumFolding2023}. Supported by an extensive literature, this principle is believed to be the universal underlying basis behind all kinetic phenomena (thermal, electrical, mechanical, chemical, etc.) and has, in fact, recently been proposed as a fourth law of thermodynamics~\cite{berettaFourthLawThermodynamics2020}. The objective of this work is to use the SEAQT framework to construct a generalized adsorption model that provides a comprehensive description of the adsorption process both near, and far from, equilibrium.

The manuscript is organized as follows. In Section \ref{sec:SEAQTframe}, the general principles of the SEAQT framework are presented with a focus on simple scenarios excluding particle number considerations. Section \ref{sec:SEAQTgrandcanonical1} extends the SEAQT equation of motion to include a system with a single particle number; this represents a system with one adsorbing specie. Section \ref{sec:SEAQTgrandcanonical2} further expands the framework to include two particle numbers, facilitating an understanding of the multicomponent adsorption process. Section \ref{sec:nonequi} provides general derivations within the SEAQT framework of specific cases of kinetic equations describing the adsorption process and a validation case demonstrating the accuracy of this approach. Section \ref{sec:equi} discusses how the SEAQT framework aligns with conventional models in limiting cases of single and multicomponent systems at equilibrium. Finally, Section \ref{sec:con} provides some conclusions for this work.

\section{SEAQT framework}
\label{sec:SEAQTframe}
The SEAQT equation of motion for an isolated simple or general quantum system can be expressed in its most general form as
\begin{equation} \label{eq:gen}
 \frac{d \hat{\rho}}{dt} = \frac{1}{i\hbar}[{\hat{\rho}}, \hat{H}]+\frac{1}{\tau \left( \hat{\rho} \right)}{\hat{D}(\hat{\rho})}
 \end{equation}
where $\hat \rho$ is the density operator, $t$ the time, $i$ the imaginary unit, $\hbar$ Planck's modified constant, $\hat{H}$ the Hamiltonian operator that represents a discrete spectrum of energy eigenvalues, $[{\hat{\rho}},\hat{H}]$ the commutator of $\hat{\rho}$ and $\hat{H}$, $\tau$ the relaxation parameter, and $\hat{D}(\hat{\rho})$ the dissipation operator. For the classical system under consideration (i.e., negligible quantum effects), both the density and the Hamiltonian operators are diagonal in the energy eigenvalue basis and, thus, commute, leading to the disappearance of the commutator or so-called symplectic term in Eq.~(\ref{eq:gen}). The dissipation term can be determined using a variational approach that identifies the steepest-entropy-ascent path in Hilbert space. This method involves ascending along the component of the gradient of the entropy perpendicular to a particular manifold. This manifold is formed by the observables for a given system that are conserved and that correspond to the generators of motion for the system. For example, the generators of motion are typically quantities like the identity operator, the Hamiltonian operator, and particle number operators~\cite{berettaSteepestEntropyAscent2014, liGeneralizedThermodynamicRelations2016,liSteepestentropyascentQuantumThermodynamic2016}. For the case of an isolated system when there is only a single identity operator, $\hat{I}$, and a single Hamiltonian operator, $\hat{H}$, the equation of motion for the $j^{th}$ energy eigenlevel is written as
\begin{equation} \label{eq:eqomot}
\frac{dp_j}{dt} = 
\frac{1}{\tau} 
\frac{
    \left|
    \begin{array}{ccc}
   p_j s_j & {p_j} & {e_jp_j} \\
    {\langle s \rangle} & {1} & {\langle e \rangle} \\
    {\langle es \rangle} & {\langle e \rangle} & {\langle e^2 \rangle} \\
    \end{array} 
    \right|
}{
    \left|
    \begin{array}{cc}
    {1} & {\langle e\rangle } \\
    {\langle e\rangle} & {\langle e^2 \rangle} \\
    \end{array} 
    \right|
}
 \end{equation}
where $\langle \cdot \rangle$ is an expectation value; $e$ the energy; $s$ the entropy; and $s_j = -\ln \left( p_j /g_j \right)$, $p_j$, and $g_j$ the entropy, occupation probability, and degeneracy, respectively, of the $j^{th}$ one-particle energy eigenlevel. Note that the determinant in the denominator is a Gram determinant, which ensures the linear independence of the generators of motion. The expectation values are expressed as
\begin{align} \label{eq:candef}
\langle e \rangle &= \sum_{j}p_je_j \\ 
\langle s \rangle &= \sum_{j}p_js_j \\ 
\langle e^2 \rangle &= \sum_{j}p_je_j^2 \\ 
\langle e s \rangle &= \sum_{j} p_je_j s_j 
\end{align}

Expanding the determinants, Eq.~(\ref{eq:eqomot}) can be reduced to 
\begin{equation}
\frac{dp_j}{dt} = \frac{1}{\tau} \Bigl( \beta\; p_j \bigl( \langle e \rangle - e_j \bigr) - p_j \bigl( \langle s \rangle - s_j \bigr) \Bigr)
\label{eq:eqomot:reduced}
\end{equation}
where $\beta = A_{es} / A_{ee}$ is the non equilibrium thermodynamic parameter that is proportional at stable equilibrium to the inverse of temperature and $A_{ee} = \langle e^2 \rangle - \langle e \rangle \langle e \rangle$ and $A_{es} = \langle e s \rangle - \langle e \rangle \langle s \rangle$ are the energy  and entropy-energy  fluctuations. Alternatively, Eq.~(\ref{eq:eqomot}) can be manipulated with elementary row operations to simplify the calculation of the determinants. The strategy to do this involves eliminating the terms above and below the ``1'' element (the second element of the second row). For instance, multiplying the second row by $p_j$ and subtracting it from the first row eliminates the second element of the first row. Also, multiplying the second row by $\langle e \rangle$, and subtracting it from the third row eliminates the second element of the third row. In this manner, Eq.~(\ref{eq:eqomot}) can be written equivalently as
\begin{equation}  \label{eq:eqomotAlt}
    \frac{dp_j}{dt} = \frac{1}{\tau} 
\frac{
    \left|
    \begin{array}{ccc}
   p_j s_j - \langle s \rangle p_j & 0 & p_j e_j - \langle e \rangle p_j\\
    {\langle s \rangle} & {1} & {\langle e \rangle} \\
    A_{es} & 0 & A_{ee} \\
    \end{array} 
    \right|
}{
    \left|
    \begin{array}{cc}
    {1} & {\langle e\rangle } \\
    {0} & A_{ee} \\
    \end{array} 
    \right|
}
\end{equation}
Expanding the determinants for Eq.~\ref{eq:eqomotAlt} is simpler than for Eq.~\ref{eq:eqomot}, but it leads to the same result: the equation of motion given by Eq.~\ref{eq:eqomot:reduced}.  Now, defining the local non-equilibrium free energy with respect to the $j^{th}$ energy eigenlevel (``local'' is used here to refer to a particular energy level rather than in a spatial sense): 
\begin{equation}
f_j = e_j - \beta^{-1} s_j \label{Eq:fj}
\end{equation}
the expectation value of the non-equilibrium free energy can be written as $\langle f \rangle = \sum_j p_j f_j$. With this expression, Eq.~(\ref{eq:eqomot:reduced}) simplifies to:
\begin{equation} \label{eq:freefluctuation1}
    \frac{dp_j}{dt} = \frac{\beta}{\tau} p_j\Bigl( \langle f \rangle   - f_j \Bigr) 
\end{equation}
The quantity $\langle f \rangle  - f_j $ is a deviation from the mean $\langle f \rangle$, and the variance of this quantity is its fluctuation. Within the SEAQT framework, this equation shows the evolution of each occupation probability is proportional to its free energy fluctuation, and at stable equilibrium, these fluctuations go to zero. 

In the limit, as the equation of motion evolves to stable equilibrium (maximum entropy) guided by the principle of SEA at every instant of time. When equilibrium is reached, the occupation probabilities stop changing, i.e.,  
\begin{equation}
   \lim_{t \rightarrow t^{ \text{se} }} \frac{dp_j}{dt} =0 \label{Eq:limitdpj/dt}
\end{equation}
and it can be shown that the non-equilibrium free energy reduces to its stable equilibrium definition. Substituting Eqs. (\ref{Eq:fj}) and (\ref{eq:freefluctuation1}) and the expression for $s_j$ into Eq.~(\ref{Eq:limitdpj/dt}) results in  
\begin{equation}
  \ln \frac{p_j}{g_j}  = -\beta_{\text{eq}} e_j + \beta_\text{eq} \langle f \rangle
\end{equation}
Taking the exponential of both sides of this last expression, the canonical distribution at stable  equilibrium is recovered, namely,
\begin{equation} \label{eq:condist}
    p_j = \frac{ g_j \exp \left( -\beta_{\text{eq}} e_j\right)}{\exp \left( -\beta_{\text{eq}} \langle f \rangle \right)} \,,
\end{equation}
where it can be shown  that the logarithm of the canonical partition function $Z = \sum_i e^{-\beta_{eq}e_i}$ is 
\begin{equation}
    \ln Z = - \beta_{\text{eq}} \langle f \rangle
\end{equation}
This last expression can be found via substitution into $\langle f \rangle$ of the expressions for the expectation values $\langle e \rangle$ and $\langle s \rangle$ and recognition that from the maximum entropy principle $\langle e \rangle = \sum_i p_ie_i = \sum_i e^{-\beta_{eq}e_i}e_i/Z = -(\partial \ln Z /\partial \beta_{eq})$.

As to $\beta_{\text{eq}}$, it can be determined from the evolution of the the rate of change of entropy. In the SEAQT framework, this rate of change depends upon the temperature, the relaxation parameter $\tau$, and the fluctuations of the free energy and entropy. The rate of change of the entropy can be derived straightforwardly using
\begin{equation}
    \frac{d\langle s \rangle}{dt} = \frac{d}{dt} \sum_j p_j s_j \,, \label{Eq:dsdt1}
\end{equation}
Substituting Eq.~(\ref{eq:freefluctuation1}) and the expression for $s_j$ given above in terms of $p_j$ and $g_j$, Eq.~(\ref{Eq:dsdt1}) reduces to 
\begin{equation}
    \frac{d\langle s \rangle}{dt} = \frac{\beta}{\tau}A_{fs} \,, 
\end{equation}
where the free energy-entropy fluctuation parameter $A_{fs} = \langle f\,s \rangle - \langle f \rangle \langle s \rangle $. At stable equilibrium, $A_{fs}$ goes to zero, and, as a consequence, the correlation of the free energy and entropy factors such that
\begin{equation}
    \langle f\,s \rangle = \langle f \rangle \langle s \rangle \,, \label{Eq:fsequil}
\end{equation} 
Now, since at stable equilibrium 
\begin{equation}
\langle f\,s \rangle =  \langle e\,s \rangle - \beta_{\text{eq}}^{-1} \langle s^2 \rangle 
\end{equation}
which results from multiplying Eq.~(\ref{Eq:fj}) by $p_j s_j$ and summing over all $j$, and 
\begin{equation}
\langle f \rangle =  \langle e \rangle - \beta_{\text{eq}}^{-1} \langle s \rangle    
\end{equation}
which results from multiplying Eq.~(\ref{Eq:fj}) by $p_j$ and summing over all $j$, Eq.~(\ref{Eq:fsequil}) can be solved for $\beta_{\text{eq}}$ such that 
\begin{equation} 
    \beta_{\text{eq}} = \left( \frac{A_{ss}}{A_{es}}\right) = \left( \frac{A_{ss}}{A_{ee}} \right)^{1/2}\,. \label{Eq:betaeq}
\end{equation}
The second equality arises from using the general definition for $\beta$ given after Equation~\ref{eq:eqomot:reduced} to replace $A_{es}$ in the first equality.  In general (i.e., at both non-equilibrium and stable equilibrium) $\beta = A_{ee}/A_{es}$ and $A_{ee} = \langle e^2 \rangle - \langle e \rangle^2$, $A_{ss} = \langle s^2 \rangle - \langle s \rangle^2$, and $A_{es} = \langle e s \rangle - \langle e \rangle \langle s \rangle$ are the energy-energy, entropy-entropy, and energy-entropy fluctuations. Equation~\ref{Eq:betaeq} is the stable equilibrium value of the intrinsic property $\beta$. Note that this last expression implies that at the absolute zero temperature ($\beta_{\text{eq}}=\infty$), there are no entropy fluctuations affecting the system. 

Now, to analyze an adsorption process, the following section presents a development of the SEAQT equation of motion that accounts for fluctuations of the grand potential, i.e., $\langle \Phi \rangle = \langle e \rangle - \beta^{-1} \langle s \rangle - \sum_i \gamma_i \beta^{-1} \langle n_i \rangle$, and its recovery at stable equilibrium. Clearly, at stable equilibrium, $\gamma_i$ is proportional to the chemical potential, $\mu_i$.

\section{The  grand canonical ensemble and kinetics of the SEAQT equation of motion}
\label{sec:SEAQTgrandcanonical1}
In order to recover the grand canonical ensemble as the stable equilibrium solution within the SEAQT framework, the dissipative operator in Eq.~(\ref{eq:gen}) is modified to include additional generators of motion, namely, a number operator, $\hat{n}$, for species $n$. The development in this section is limited to a single species; the next section considers two species. These developments are easily extended to any number of species. 

For the analysis of adsorption kinetics, it is assumed that the observable, $\langle n \rangle$, of the number operator, $\hat{n}$, is conserved throughout the kinetic process and represents the total number (of molecules) of the adsorbing specie. It can be present as an adsorbed molecule, $\langle n \rangle^{ad}$, or as a free, unadsorbed molecule, $\langle n \rangle^{un}$. Thus, $\langle n \rangle = \langle n \rangle^{ad} + \langle n \rangle^{un}$. The generators of motion for this single species adsorption problem are then $\hat{I}$, $\hat{H}$, and $\hat{n}$. Again, as before, the symplectic component of the dynamics is zero, leading to the following equation of motion:
\begin{equation} \label{eq:geneq}
\frac{dp_j}{dt} = 
\frac{1}{\tau} 
\frac{
    \left|
    \begin{array}{cccc}
   p_j s_j & {p_j} & {e_jp_j}  & ({n_j^{ad}+n_j^{un}) p_j}\\
    {\langle s \rangle} & {1} & {\langle e \rangle} & \langle n \rangle \\
    {\langle es \rangle} & {\langle e \rangle} & {\langle e^2 \rangle} & \langle e n \rangle\\
    \langle ns  \rangle & \langle n \rangle & \langle e n \rangle & \langle n^2 \rangle 
    \end{array} 
    \right|
}{
    \left|
    \begin{array}{ccc}
    {1} & {\langle e \rangle}  & \langle n \rangle \\
    {\langle e \rangle} & {\langle e^2 \rangle} &  {\langle e n \rangle} \\
    \langle n \rangle & {\langle e n \rangle} &  {\langle n^2 \rangle} \\
    \end{array} 
    \right|}
 \end{equation}
where 
\begin{align}
\langle en \rangle &= \sum_{j}p_je_j n_j\\ 
\langle ns \rangle &=  \sum_{j}p_j n_j s_j \\ 
\langle n \rangle &= \sum_{j}p_j n_j \\ 
\langle n^2 \rangle &= \sum_{j} p_j n_j^2
 \end{align}
and again $p_j$ is the occupation probability of the $j^{th}$ one-particle energy eigenlevel, $n_j= n_j^{ad}+n_j^{un}$ is the particle occupation number of species $n$ of the $j^{th}$ one-particle energy eigenlevel, and $\tau$ is the relaxation parameter that determines the velocity of the dissipative dynamics. The Gram determinant given in the denominator can be interpreted as a more complex variation of the energy fluctuation, $A_{ee}$, given by the Gram determinant in Eq.~(\ref{eq:eqomot}), i.e., one that involves not only the energy but the number of species particles as well. 

This formulation, recognized earlier by Beretta~\cite{berettaNonlinearModelDynamics2006}, stresses that the equation of motion can be written as a generalized fluctuation-dissipation formulation in terms of the covariance $\langle \Delta  A \Delta B \rangle = \langle A B \rangle - \langle A \rangle \langle B \rangle$ for any extensive properties $A$ and $B$. In the same manner used to arrive at Eq.~(\ref{eq:eqomotAlt}), elements above and below the ``1'' element (second element of the second row) are eliminated through elementary row operations and the fluctuating nature of the determinant becomes explicit. Specifically, by multiplying the second element of the second row by $p_j$ and subtracting the result from the first row, the element above the second element of the second row is eliminated. Similarly, by multiplying the second row by $\langle e \rangle$ and subtracting it from the third row, the second element of the third row is eliminated. Repeating this process for the fourth row, Eq.~(\ref{eq:candef}) is rewritten as
\begin{equation} \label{eq:geneqf}
    \frac{dp_j}{dt} = 
\frac{1}{\tau} 
\frac{
    \left|
    \begin{array}{cccc}
   p_j s_j - \langle s \rangle p_j & {0} & p_j e_j - \langle e \rangle p_j & p_j n_j - \langle n \rangle p_j\\
    {\langle s \rangle} & {1} & {\langle e \rangle} & \langle n \rangle \\
    A_{es} & {0} & A_{ee} & A_{en}\\
    A_{ns}   & 0 & A_{en} & A_{nn} 
    \end{array} 
    \right|
}{
    \left|
    \begin{array}{ccc}
    {1} & {\langle e \rangle}  & \langle n \rangle \\
    {0} & A_{ee} &  A_{en} \\
    0 &  A_{en} &  A_{nn} \\
    \end{array} 
    \right|}
\end{equation}
where $A_{en} = \langle e\, n \rangle - \langle e \rangle \langle n \rangle$, $A_{ns} = \langle n\,s \rangle - \langle n \rangle \langle s \rangle$, and $A_{nn} = \langle n^2 \rangle - \langle n \rangle \langle n \rangle$ are the energy-particle number, particle number-entropy, and particle number fluctuations.

Expanding the determinants, Eq.~(\ref{eq:geneqf}) can be rewritten in the following more compact form: 
 \begin{equation}
\frac{dp_j}{dt}=-\frac{1}{\tau}\Bigl(- p_j \ln \frac{p_j}{g_j} + \beta p_j e_j - \gamma p_j n_j -\beta \langle \Phi\rangle  p_j\Bigr) \label{Eq:eomonen}
 \end{equation}
and the expectation of the grand potential, $\langle \Phi \rangle$, is defined  as
\begin{align}
\langle \Phi  \rangle =  \langle e \rangle - \beta^{-1} \langle s \rangle - \beta^{-1}\gamma \langle n \rangle \,,
\end{align}
where the non-equilibrium intensive properties, $\beta$ and $\gamma$, are functions of the energy-particle number, energy-entropy, energy, particle number-entropy, and particle number fluctuations $A_{en}$, $A_{es}$, $A_{ee}$, $A_{ns}$, and $A_{nn}$, respectively, such that the intensive properties can be written as the ratio of the minors
\begin{align}
\beta &= \frac{M_e}{M_s} \,,\quad \gamma= \frac{M_n}{M_s} \label{Eq:betaIII}
\end{align}
where the respective minors are
 \begin{align}
M_e &= \det \left| \begin{matrix}
    A_{es} & A_{en} \\
    A_{ns} & A_{nn} 
\end{matrix} \right| \,, \\[3mm]
M_s &= \det \left| \begin{matrix}
A_{ee} & A_{en} \\
A_{en} & A_{nn} 
\end{matrix}\right|  \,, \\[3mm]
M_n &= \det \left|\begin{matrix}
A_{es} & A_{ee} \\
A_{ns} & A_{en} 
\end{matrix}
\right| \,.
\end{align}
At stable equilibrium, these non-equilibrium intensive properties match the inverse temperature,  $\beta =1/k_B T$, and chemical potential, $\mu =\gamma \, \beta^{-1}$, respectively, so that the non-equilibrium grand potential $\langle \Phi \rangle $ matches its stable equilibrium definition as well. Note that $k_B$ is Boltzmann's constant. 

Now, the local grand potential for the $j^{th}$ one-particle eigenlevel is written as 
\begin{equation}
    \Phi_j = e_j -\beta^{-1} s_j -\gamma \,\beta^{-1} n_j
\end{equation}
Using this last expression, Eq.~(\ref{Eq:eomonen}) can be written as deviations of the grand potential from the mean, i.e.,
\begin{equation}
\frac{dp_j}{dt}=\frac{\beta}{\tau} p_j \bigl( \langle \Phi \rangle - \Phi_j \bigr) \,,
\end{equation}
which allows one to interpret the time evolution of the probability distribution as the consequence of the fluctuations of the grand potential. At stable equilibrium, the dissipation ceases and the time derivative of the probability, $p_j$, goes to zero resulting in
\begin{equation}
    \ln \frac{p_j}{g_j}  = \beta_{\text{eq}} \langle \Phi \rangle  - \beta_{\text{eq}} \left(  e_j-\gamma_{\text{eq}} \beta_{\text{eq}}^{-1} n_j\right)
\end{equation}
where $\beta_{\text{eq}}$ is determined by Eq.~(\ref{Eq:betaIII}). Solving for $p_j$ and using the fact that $\gamma_{\text{eq}} \,\beta^{-1}_{\text{eq}} = \mu$, the grand canonical probability distribution is recovered such that
 \begin{equation} \label{eq:grancondist}
p_j = \frac {g_j \exp \bigl( {-\beta_{\text{eq}} \left( e_j - \mu n_j\right)}\bigr)}{\exp \bigl( -\beta_{\text{eq}}\langle \Phi \rangle \bigr)}
 \end{equation}
As expected and as can be demonstrated in a manner similar to what was done for Eq.~(\ref{eq:condist}), the logarithm of the grand canonical partition function, $\mathcal{Z}$, is $$\ln \mathcal{Z} = -\beta_{\text{eq}} \langle \Phi \rangle$$ where $\langle \Phi \rangle$ matches with the equilibrium grand potential and $\mathcal{Z} = \sum_i e^{-\beta_{eq}(e_i-\mu_{eq} n_i})$ from the maximum entropy principle. 

Of course, the SEAQT framework not only predicts the intensive properties of temperature and chemical potential at stable equilibrium consistent with those of equilibrium thermodynamics but as well their non-equilibrium counterparts via the fluctuations even far from equilibrium. Indeed, this fluctuation formulation is compatible with the minimization of the free energy at equilibrium as shown in Appendix \ref{sec:fluctuation}.

\section{Multicomponent adsorption and interaction with a thermal reservoir}
\label{sec:SEAQTgrandcanonical2}
For a multicomponent system, two or more number operators as generators of motion are required. For simplicity, only two species and hence two number operators are considered here. In addition, using the hypoequilibrium concept~\cite{liSteepestentropyascentQuantumThermodynamic2016,liGeneralizedThermodynamicRelations2016}, a 2$^{nd}$-order hypoequilibrium description is assumed to treat a system (S) interacting with a thermal reservoir (R). 
\begin{widetext}
\begin{equation} \label{eq:seaqtnumber}
\frac{dp_j}{dt} = 
\frac{1}{\tau} 
\frac{
    \left|
    \begin{array}{cccccc}
   {p_j s_j} & {p_j} & {0} & {e_jp_j} & {(n_j^{ad}+n_j^{un})p_j} & {(m_j^{ad}+m_j^{un})p_j} \\
   
    {\langle s \rangle_S} & {P_S} & {0} & {\langle e \rangle}_S &  {\langle n \rangle}_S &  {\langle m \rangle}_S \\
    
    {\langle s \rangle_R} & {0} & {P_R} & {\langle e \rangle}_R &  {\langle n \rangle_R } &  {\langle m \rangle_R} \\
    
    {\langle es \rangle_S + \langle es \rangle_R } & { \langle e \rangle}_S & {\langle e \rangle}_R & {\langle e^2 \rangle_S+\langle e^2 \rangle_R} &  {\langle ne \rangle}_S+{\langle ne \rangle}_R &  {\langle me \rangle}_S+{\langle me \rangle}_R \\
    
    {\langle ns \rangle_S+\langle ns \rangle_R} & {\langle n \rangle}_S & {\langle n \rangle_R} & {\langle ne\rangle}_S+{\langle ne\rangle}_R &  {\langle n^2 \rangle_S+\langle n^2 \rangle_R} &  {\langle nm \rangle_S+\langle nm \rangle_R} \\  
    
      {\langle ms \rangle_S+\langle ms \rangle_R} & {\langle m \rangle}_S & {\langle m \rangle_R} & {\langle me\rangle}_S+{\langle me\rangle}_R &  {\langle nm \rangle_S+\langle nm \rangle_R} &{\langle m^2 \rangle_S+\langle m^2 \rangle_R}  \\  
    \end{array} 
    \right|
}{
    \left|
    \begin{array}{ccccc}
    {P_S} & {0} &  { \langle e \rangle}_S & {\langle n \rangle}_S &  {\langle m \rangle}_S  \\
    
    {0} & {P_R} & {\langle e \rangle}_R & {\langle n \rangle}_R &  {\langle m \rangle}_R\\
   
   { \langle e \rangle}_S & {\langle e \rangle}_R &   {\langle e^2 \rangle_S + \langle e^2 \rangle_R }  &  {\langle ne \rangle}_S+{\langle ne \rangle}_R &  {\langle me \rangle}_S+{\langle me \rangle}_R \\ 
    
  {\langle n \rangle}_S & {\langle n \rangle}_R & {\langle ne\rangle}_S+{\langle ne\rangle}_R &  {\langle n^2 \rangle_S}+{\langle n^2 \rangle_R} &  {\langle nm \rangle_S }+{\langle nm \rangle_R} \\  
    
       {\langle m \rangle}_S & {\langle m \rangle}_R & {\langle me\rangle}_S+{\langle me\rangle}_R &  {\langle nm \rangle_S} +{\langle nm \rangle_R} &{\langle m^2 \rangle_S}+{\langle m^2 \rangle_R}
    
    \end{array} 
    \right|
}
 \end{equation}
 
 \end{widetext}
The system and the reservoir comprise a composite system for which there are five generators of motion, namely, the Hamiltonian operator, $\hat{H}$, for the composite; an identity operator for the system, $\hat{I}_S$, and one for the reservoir, $\hat{I}_R$; and a particle number operator, $\hat{n}$, for the first specie and another one, $\hat{m}$, for the second specie. The observables associated with each of these operators (i.e., $\{\tilde{C}(\textbf{\textit{p}})\}=\{\tilde{H},\tilde{n},\tilde{m},\tilde{I}_S,\tilde{I}_R\}$) are conserved by the equation of motion, Eq.~(\ref{eq:seaqtnumber}), which is the equation of motion for the system ($S$). An equation of motion for the reservoir ($R$) is not needed since it is a thermal reservoir whose state remains unchanged, i.e., any changes to the total energy and entropy of the reservoir are by definition negligible and, thus, do not affect its state. Therefore, the reservoir expectation values associated with the energy and entropy appearing in Eq.~(\ref{eq:seaqtnumber}) only represent the energy and entropy added to or subtracted from the reservoir. 

Now, each of the last five columns in the determinant of the numerator of Eq.~(\ref{eq:seaqtnumber}) corresponds to one of the generators of motion, while the first column corresponds to the entropy operator, $\hat{S}$. Normalizing the expectation values associated with each observable such that for $i=\{S, R\}$, $\langle I \rangle_i = P_i \langle \grave{I} \rangle_i$ where $I=\{e, s, n, m, es, ns, ne, ms, me, e^2, s^2, n^2, m^2\}$, Eq.~(\ref{eq:seaqtnumber}) can be reduced by eliminating the terms above and below the element $P_S$ and $P_R$ through row operations. For instance, by multiplying the second row by $p_j$ and subtracting the result from the first row, the second element of the first column is eliminated. This procedure is repeated to eliminate the elements $\langle e \rangle_S$, $\langle n \rangle_S$, and $\langle m \rangle_S$. In the same manner, the elements below $P_R$ in the third column are eliminated by row operations. In this case, the third row can be multiplied by $\langle e \rangle_R/P_j$, and the result subtracted from the fourth row. Repeating the process for the fifth and sixth rows eliminates the elements $\langle n \rangle_R$ and $\langle m \rangle_R$. In this manner, and in the limit of $P_R \gg P_S$, Eq.~(\ref{eq:seaqtnumber}) can be compactly written as 
 \begin{align} \label{eq:grand}
\frac{dp_j}{dt}&=\frac{1}{\tau}\Bigl( p_j s_j + p_j\bigl( \beta_R \langle e \rangle_S -  \langle s \rangle_S - 
\gamma_{n}\langle n \rangle_S   \\ \nonumber
&- \gamma_{m} \langle m \rangle_S  \bigr)-p_j\bigl( \beta_R {e}_j  -  \gamma_{n}n_j - \gamma_{m}m_j \bigr) \Bigr) \,.
 \end{align}

The species are comprised of both adsorbed and unadsorbed species so that $\langle n \rangle_S = \langle n \rangle_{ad} + \langle n \rangle_{un}$ and $\langle m \rangle_S = \langle m \rangle_{ad} + \langle m \rangle_{un}$ and the occupation particle numbers are $n_j= n_j^{ad}+n_j^{un}$ and $m_j = m_j^{ad}+m_j^{un}$ for each $j^{th}$ energy eigenlevel. In addition, the intensive properties are defined in terms of the generalized fluctuations such that
\begin{equation} \label{MultiCompMinors}
\beta_R=\frac{M_e}{M_s}\,,\;\;\gamma_{n}=\frac{M_n}{M_s} \,,\;\;\gamma_{m}=\frac{M_m}{M_s}
\end{equation}
where $\beta_R=1/k_B\,T_R$ is the inverse of the product of Boltzmann's constant, $k_B$, and the reservoir temperature, $T_R$; $\gamma_{n}$ is the dimensionless chemical potential for species $n$; and $\gamma_{m}$ the dimensionless chemical potential for species $m$. The generalized fluctuation components  are defined via the following minors of the dissipative matrix
\begin{equation}
M_s = \det 
    \left|
    \begin{array}{ccc}
   A_{ee} & A_{en} & A_{em} \\
     A_{en} & A_{nn} & A_{nm} \\
     A_{em} & A_{nm} & A_{mm} \\
    \end{array} 
    \right| \,,
 \end{equation}
 
\begin{equation}
M_e = \det 
    \left|
    \begin{array}{ccc}
   A_{es} & A_{en} & {A_{em}} \\
     A_{ns} & {A_{nn}} & {A_{nm}} \\
     {A_{ms}} & {A_{nm}} & {A_{mm}} \\
    \end{array} 
    \right| \,,
 \end{equation}

\begin{equation}
M_n = \det 
    \left|
    \begin{array}{ccc}
   {A_{es}} & {A_{ee}} & {A_{em}} \\
     {A_{ms}} & {A_{em}} & {A_{mm}} \\
          {A_{ns}} & {A_{en}} & {A_{nm}}
    \end{array} 
    \right| \,,
 \end{equation}
and
\begin{equation}
M_m = \det 
    \left|
    \begin{array}{ccc}
   A_{es} & A_{ee} & A_{en} \\
     A_{ns} & A_{en} & A_{nn} \\
     A_{ms} & A_{em} & A_{mn} \\
    \end{array} 
    \right| \,.
 \end{equation}
Each binary fluctuation is given by $A_{IJ} = \langle IJ \rangle - \langle I \rangle  \langle J \rangle$ for $I,J={\{e,s,n,m\}}$. Note that for the case when $I=n$ equals some constant $k_n$ and $I=m$  equals some constant $k_m$,  the correlators $\langle IJ \rangle$ and $\langle JI \rangle$ factor as $\langle I \rangle  \langle J \rangle$, and the fluctuations $A_{ns}$ as well as $A_{en}$ go to zero. As a consequence, the dimensionless chemical potentials, $\gamma_n$ and $\gamma_m$, vanish, recovering the formulation of the equation of motion without the number operators as generators of motion. 

Now, consider the grand potential 
\begin{equation}\label{eq:grandpot}
    \Phi_j = e_j- \beta_R^{-1}  s_j - \mu_{R_n}\,  n_j- \mu_{R_m}\, m_j
\end{equation}
where $\mu_{R_n} = M_{n}/ M_{e}$ and $\mu_{R_m} = M_m/ M_{e}$ are the usual chemical potentials for species $n$ and $m$. The associated expectation value for the non-equilibrium grand potential is 
\begin{equation}
    \langle \Phi \rangle = \sum_j p_j \Phi_j.
\end{equation}
Using this last expression, Eq.~(\ref{eq:grand}) is rewritten as
\begin{equation}\label{eq:grandmod}
   \frac{d p_j}{dt}=\frac{ \beta_R}{\tau} p_j \bigl( \langle \Phi \rangle - \Phi_j   \bigr),
\end{equation}
which indicates that the evolution of the probabilities via the SEAQT equation of motion is related to the deviation of the grand potential from its mean value (the variance of which is the fluctuation of the grand potential).  In addition, the stable equilibrium solution for the probability distribution predicted by the SEAQT equation of motion takes the form
\begin{equation} \label{eq:grancandisteq}
    p_j = \frac{g_j \exp \Bigl( - \beta_R \bigl( e_j - \left(\mu_{R_n} n_j+\mu_{R_m} m_j\right)\bigr)  \Bigr)}{\exp \bigl( - \beta_R \langle \Phi \rangle \bigr)}
\end{equation}
where $- \beta_R \langle \Phi \rangle = \ln \mathcal{Z}$ is the natural logarithm of the grand canonical partition function, $\mathcal{Z}$, which from the maximum entropy principle, is $\mathcal{Z} = \sum_i e^{-\beta_R (e_i-(\mu_{R_n} n_i + \mu_{R_m} m_i))}$.
\section{Non-equilibrium dynamics}
\label{sec:nonequi}
Consider now the evolution to stable equilibrium of the  expectation value of the particle number defined by
\begin{equation}
\langle n \rangle_{ad} = \sum_j p_j(t)\, n_j^{ad}
\end{equation}
where $p_j$ is the occupation probability of the $j^{th}$ energy eigenlevel and $n_j^{ad}$ is the particle occupation number for this eigenlevel for the adsorbed species, which does not depend upon time. The occupation numbers can be obtained, for example, from a Monte Carlo simulation such as the non-Markovian Replica-Exchange Wang-Landau algorithm~\cite{wangEfficientMultipleRangeRandom2001,wangDeterminingDensityStates2001,vogelGenericHierarchicalFramework2013,liNewParadigmPetascale2014}. The dynamics of the particle number under the SEAQT ansatz is then written as
\begin{equation}\label{eq:particleeom}
\frac{d \langle n \rangle_{ad}}{dt} = \sum_j \frac{dp_j}{dt} n_j^{ad}
\end{equation}
where direct substitution of Eq.~(\ref{eq:grandmod}) results in 
\begin{equation}
    \frac{d\langle n \rangle_{ad}}{dt} = \frac{ \beta_R}{\tau}\sum_j    \left( p_j n_j^{ad} \Phi_j - p_j n_j^{ad} \langle \Phi \rangle  \right)  
\end{equation}
Using the expectation value
\begin{align}
\langle  \Phi n \rangle_{ad} &= \sum_{j} p_j \Phi_j n_j^{ad}\,, 
\end{align}
Eq.~(\ref{eq:particleeom}) can be rewritten as
\begin{equation}\label{eq:kinetic}
    \frac{d\langle n \rangle_{ad}}{dt} = - \frac{ \beta_R}{\tau} \left( \langle \Phi n \rangle_{ad} -\langle \Phi \rangle \langle n \rangle_{ad} \right)
\end{equation}
This transport equation accounts  for the time evolution of the adsorbed particle number arising from the grand potential-particle number fluctuations.  This form of the transport equation is closely related to the fluctuation-dissipation relation of reference~\cite{crooksMeasuringThermodynamicLength2007}. 

Now, consider the simplest case where the  the grand potential and the number operator are not correlated, i.e., $\langle \Phi n \rangle_{ad} = 0$. Using this condition and Eq.~(\ref{eq:grandpot}) and assuming a single absorbed species results in
\begin{equation} \label{eq:cond}
    \beta_R \langle e n \rangle_{ad} -  \mu_n  \beta_R \langle n^2 \rangle_{ad} = \langle n s \rangle_{ad} 
\end{equation}
For ideal classical gases, the chemical potential is negative in the high-temperature regime since the average distance between particles is much greater than the particle wavelength~\cite{sevillaChemicalPotentialInteracting2012}. Thus, both terms on the left-hand side of Eq.~(\ref{eq:cond}) are positive provided the eigen-energies of the gas are positive, i.e., $e_i >0$. This is the case since no intermolecular forces are in play. As a result, the entropy and particle number are positively correlated so that any increase in the particle number results in an increase in the entropy. In the case of a real classical gas, the chemical potential is positive in the high-temperature regime~\cite{sevillaChemicalPotentialInteracting2012}. Thus, when the particle interactions involve attractive forces, $e_i > 0$ and provided these interactions are sufficiently strong, the first term on the left of Eq.~(\ref{eq:cond}) dominates the second term. As a consequence, the particle number and entropy are again positively  correlated. In either case, one obtains either a first order or a second order kinetics, since Eq.~(\ref{eq:kinetic}) reduces to
\begin{equation} \label{eq:kin_12}
    \frac{d\langle n \rangle_{ad}}{dt} =  \frac{ \beta_R}{\tau} \left( \langle f \rangle \langle n \rangle_{ad} -  \mu_{R_n} \left(\langle n \rangle_{ad}\right)^2 \right) \,.
\end{equation}
 This equation has a general solution for an initial number of particles $\langle n  {\scriptstyle (t=0)} \rangle_{ad} = \langle n_0\rangle_{ad}$ of 
\begin{eqnarray}
\label{Eq:gensoleq41}
    q^{neq} &=& \frac{\langle n \rangle_{ad}}{\langle n_0 \rangle_{ad}} \nonumber \\ &=&  \frac{\langle f \rangle\exp\bigl( \beta_R \langle f \rangle t \bigr)  }{\langle f \rangle + \mu_{R_n} \langle n_0 \rangle_{ad}  \exp\bigl( \beta_R \langle f \rangle t \bigr)- \mu_{R_n} \langle n_0 \rangle_{ad} }
\end{eqnarray}
 which has an equilibrium solution given by 
 \begin{equation}
     \langle n \rangle^{\text{eq}}_{ad}=  \frac{\langle f \rangle}{\mu_{R_n}}
 \end{equation}
provided $\langle f \rangle t > 0$. 

\begin{figure}[htb]
   \centering
   \includegraphics[scale=0.33]{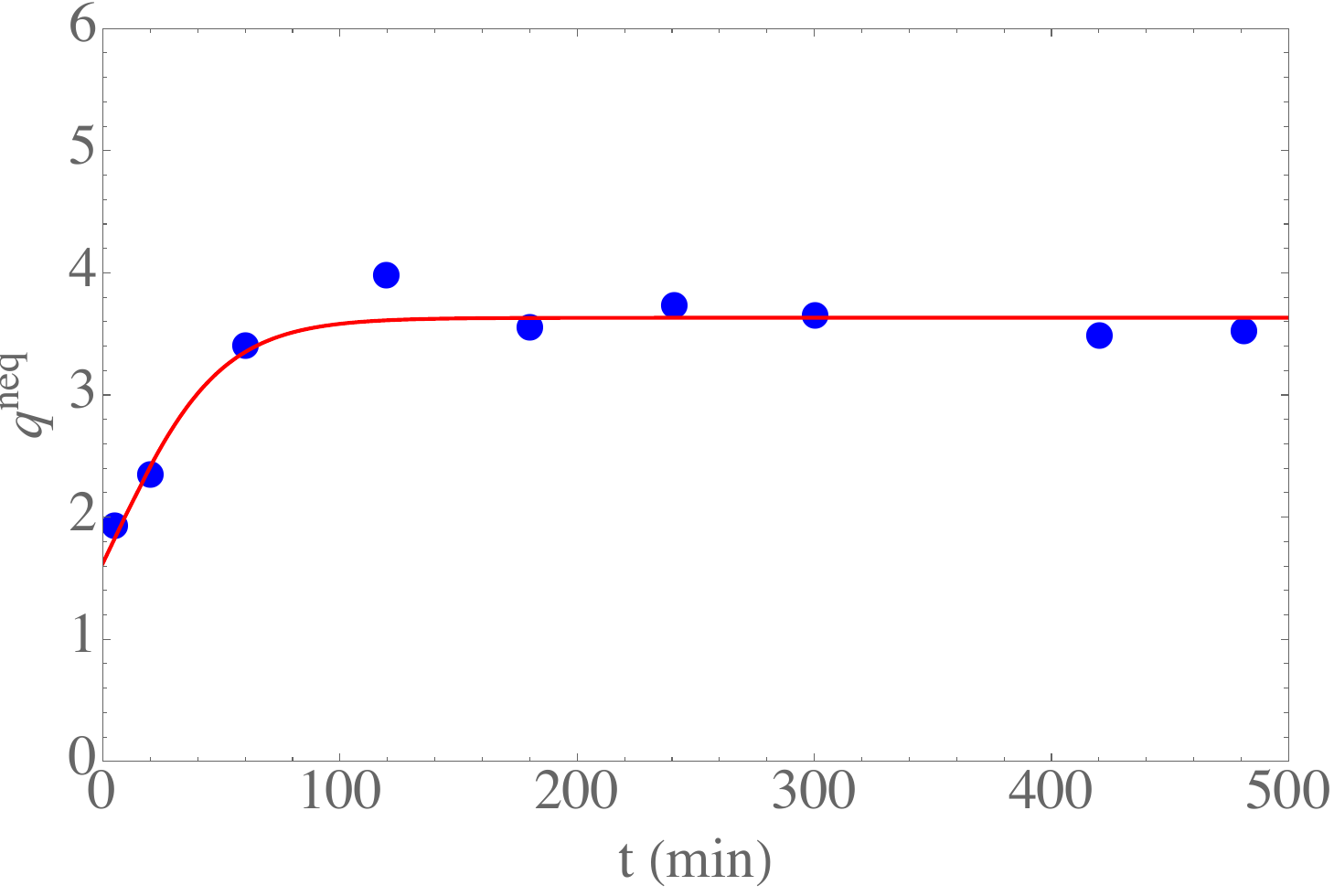} 
   \caption{ \label{fig:kin12} The kinetics of adsorption for As (V) onto magnetic graphene oxide. The blue dots represent the experimental data reported in~\cite{yuEnhancedRemovalPerformance2015}, and the red line the prediction resulting from the kinetic path determined by the SEAQT equation of motion. }
\end{figure}

In order to compare the SEAQT predictions with experimental data, consider the case of As(V) adsorption onto magnetic graphene oxide~\cite{yuEnhancedRemovalPerformance2015}. In Fig.~\ref{fig:kin12}, the experimental data reported in~\cite{yuEnhancedRemovalPerformance2015} is compared with the kinetic predictions of the SEAQT framework given by Eq.~(\ref{Eq:gensoleq41}). In this plot, the best-fit requires $\mu_{R_n} = 1.61$, $\beta_R = 0.0076$, $\langle n_0 \rangle = 1.62266$, and $\langle f \rangle = 5.85$, resulting in a correlation coefficient of $0.9980$. $\beta_R$ and $\langle f \rangle$ are in arbitrary units of energy and $\mu_{R_n}$ in arbitrary units of energy per number of particles. Of course, in an actual simulation based upon an energy eigenstructure developed for the system tested in~\cite{yuEnhancedRemovalPerformance2015}, the free energy would vary in such a way as to provide a similar fit of the experimental data. In fact, this is done in an accompanying paper~\cite{saldana-roblesSteepestentropyascentFrameworkPredicting2024}. Furthermore, in contrast to the usual kinetic models, which require an {\it a priori} assumption of a kinetic coefficient and a constitutive expression, the predictions made by the SEAQT framework are based upon a unique non-equilibrium thermodynamic path dictated by choosing the SEA direction at each instant of time.

Now consider the case where the non-equilibrium free energy is zero. A zero non-equilibrium free energy in this context would imply a particular balance between the energetic contributions and the entropic contributions across the system's non-equilibrium states where the contributions from the energy eigenlevels and entropy counteract each other exactly when weighted by the probabilities of a given state probability distribution. In this case, Eq.~(\ref{eq:kin_12}) reduces to 
\begin{equation} \label{eq:kin2}
      \frac{d\langle n \rangle_{ad}}{dt} =  -\frac{ \beta_R}{\tau}  \mu_{R_n} \left(\langle n \rangle_{ad}\right)^2 \,
\end{equation}
which, of course, represents a second-order kinetics. Notice that the rate of change of the particle number increases as the relaxation parameter $\tau$ becomes smaller. In contrast with the usual psuedo-second-order kinetics used in the literature, i.e., the second order velocity constant, $k_2$, the parameters obtained from the SEAQT framework are purely thermodynamic with $k_2$ expressed in terms of intrinsic thermodynamic properties such that
\begin{align}
k_2 &= \frac{\beta_R \, \mu_{\text{\tiny R}_n}}{\tau}
\end{align}
In general, $k_2$ only determines the velocity at which the rate of change occurs and must be fixed by experimental data. 

For the case when the chemical potential is zero, the kinetics is first order so that
\begin{equation} \label{eq:kin1}
      \frac{d\langle n \rangle_{ad}}{dt} =  \frac{ \beta_R}{\tau} \langle f \rangle \langle n \rangle_{ad} \,.
\end{equation}
In this case, the first-order velocity constant, $k_1$, is given by
\begin{equation}
k_1 = \frac{\beta_R}{\tau} \langle f \rangle
\end{equation}
and depends upon the intensive variable $\beta_R$ as well as the average free energy. Notice that both kinetic velocity constants are inversely proportional to the relaxation parameter, $\tau$, indicating that the longer the relaxation time towards equilibrium, the smaller the kinetic velocity constant. This result is as expected and allows this free parameter in the SEAQT framework to be fixed via experimental data. 

Remarkably, the first- and second-order kinetics are just special cases of the SEAQT kinetic equation given by Eq.~(\ref{eq:kinetic}), and indeed this equation represents a completely general kinetics towards stable equilibrium.

\section{Stable Equilibrium}
\label{sec:equi}
\subsection{Single component Langmuir equation}

As was shown in the previous section, the SEAQT equation of motion, Eq.~(\ref{eq:grandmod}), reduces to Eq.~(\ref{eq:grancandisteq}) at stable equilbirum where the denominator of the latter is the partition function $\mathcal{Z}=e^{-\beta_R \langle \Phi \rangle}$ and $\langle \Phi \rangle$, which is the grand potential, is given for a system with a single species $n$ by
\begin{equation}
\langle \Phi \rangle = \langle e \rangle - \beta^{-1}_R \langle s \rangle - \mu_{R_n} \langle n \rangle
\end{equation}
Furthermore, as previously shown, $\mathcal{Z} = e^{-\beta_R \langle \Phi \rangle} = \sum_i e^{-\beta_R (e_i-\mu_{R_n} n_i)}$ where the argument of the second exponential has been reduced to that for a single species. The average particle number adsorbed at stable equilibrium then becomes  
\begin{equation}
\langle n \rangle_{ad}^{eq} = \sum_j n_j^{ad} p_j^{eq}
\end{equation}
where  $p_j^{eq}$ is the stable equilibrium probability (i.e., Eq.~(\ref{eq:grancandisteq})) predicted by the SEAQT equation of motion. 

To obtain the equivalent Langmuir equation result, only the adsorbed species part of the grand canonical partition function at stable equilibrium is used such that
\begin{equation} \label{LangAdsNo}
\mathcal{Z}^{ad} = \sum_{j=1}^{N} g_j \exp \left( -\beta_R e_j + \gamma_n n_j^{ad} \right)
\end{equation}
The Langmuir isotherm model can then be found by assuming that only the first two terms of the grand canonical partition function are non-negligible. This condition, physically, is related to cases where the degeneracies at low energies are larger than the degeneracies at more elevated energies and the number of bounds represented by $n_j^{ad}$, are only relevant for the lower energies. In this way, the adsorption is compatible with a monolayer adsorption and $\mathcal{Z}^{ad}$ takes the form 
\begin{align}
\mathcal{Z}^{ad} &= g_0 \exp(-\beta_R e_0 + \gamma_n n_0^{ad} ) \\ \nonumber
&+ g_1 \exp(-\beta_R e_1 + \gamma_n n_1^{ad}) 
\end{align}

The Langmuir isotherm model assumes that adsorption occurs only in monolayers and that each adsorption site does not interact with adjacent adsorption sites. In this case, the lowest energy level can be set to zero so that the dimensionless chemical potential is zero. As a result,
\begin{equation}
\mathcal{Z}^{ad} = 1+ g_1 \exp(-\beta_R e_1 + \gamma_n n_1^{ad} )  \,,
\end{equation}
where the degeneracy for the zeroth energy eigenlevel is 1. This partition function can then be written in terms of the absolute activity for  a collection of particles, $ \xi_n^{ad} = \exp \left( \gamma_n n_1^{ad} \right)$, such that
\begin{equation}
\mathcal{Z}^{ad} = 1+ g_1 \exp \left( -\beta_R  e_1 \right) \xi_n^{ad} 
\end{equation}

Now, if there are $M$ such collections of particles available for adsorption, then the total partition function is given by
\begin{equation}
\mathcal{Z}_T^{ad} = \prod_i^M\mathcal{Z}_i^{ad} 
\end{equation}
At stable equilibrium, the total number of particles adsorbed can then be found from
\begin{equation}
\langle n \rangle^{eq}_{ad} = \xi_n^{ad} \left( \frac{\partial \ln \mathcal{Z}_T^{ad}}{\partial \xi_n^{ad}} \right)
\end{equation}
which results in
\begin{equation} \label{eq:ads}
\frac{\langle n \rangle^{eq}_{ad}}{M} = \frac{g_1 \xi_n^{ad}  \exp \left( - \beta_R  e_1 \right)}{1+ g_1\xi_n^{ad}  \exp \left( - \beta_R  e_1 \right)}
\end{equation}
This last equation is in effect the well-known Langmuir isotherm model where $C_1 = g_1 \exp \left( - \beta_R  e_1 \right)$ is proportional to the concentration and $\xi_n^{ad} = b$ is proportional to the Langmuir constant, which is a measure of how strongly a given particle interacts with a given absorption site. Defining the ratio $\langle n \rangle_{eq}^{ad} / M \equiv q_1$, Eq.~(\ref{eq:ads}) can be recast in the standard Langmuir form as
\begin{equation}
q_1 = \frac{b\, C_1}{1+ b\, C_1} \,,
\end{equation}
where $q_1$ is interpreted as the maximum adsorption capacity. This interpretation is indeed consistent with monolayer adsorption since a higher number of particles available for adsorption increases the likelihood of a particle being retained on the surface of the adsorbent. Thus, the SEAQT solution at stable equilibrium is compatible with the standard Langmuir model.
\begin{figure}[htbp]
   \centering
   \includegraphics[scale=0.35]{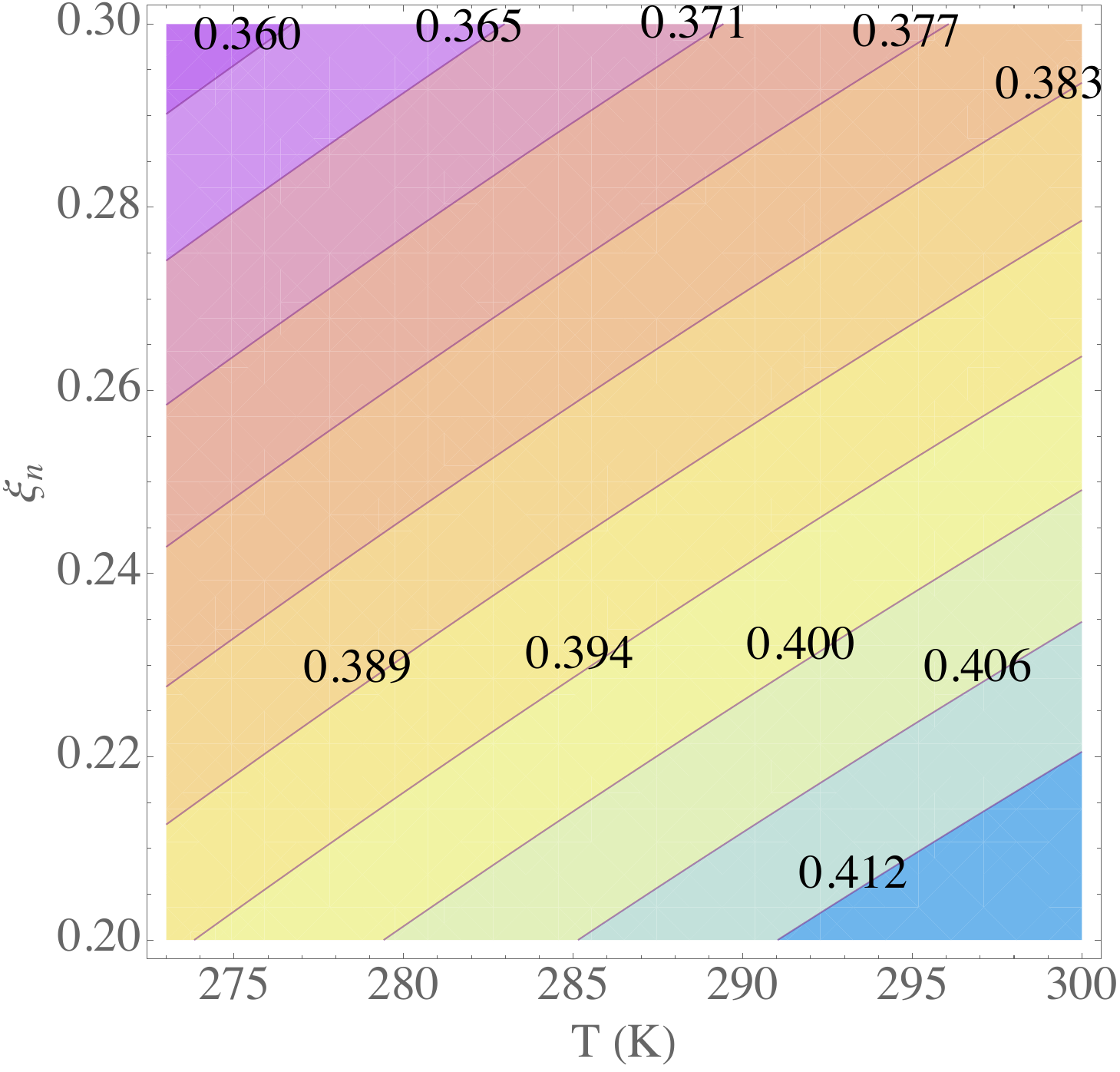} 
   \caption{ \label{fig:fig2} Variation of the dimensionless Langmuir constant as a function of temperature and the product of $b\,C_1$. Contour curves of  $b\,C_1$ are indicated. }
\end{figure}

Fig.~\ref{fig:fig2} shows the dependence of the dimensionless Langmuir constant, $\xi_n^{ad}=b$ as a function of both the temperature and $bC_1$. As seen, increasing the temperature increases $bC_1$, while increasing the chemical potential decreases it. As a consequence, the maximum adsorption capacity, $q_1$ increases with temperature but at a given temperature decreases with increases in the chemical potential.

\begin{figure}
    \centering
    \includegraphics[scale=0.35]{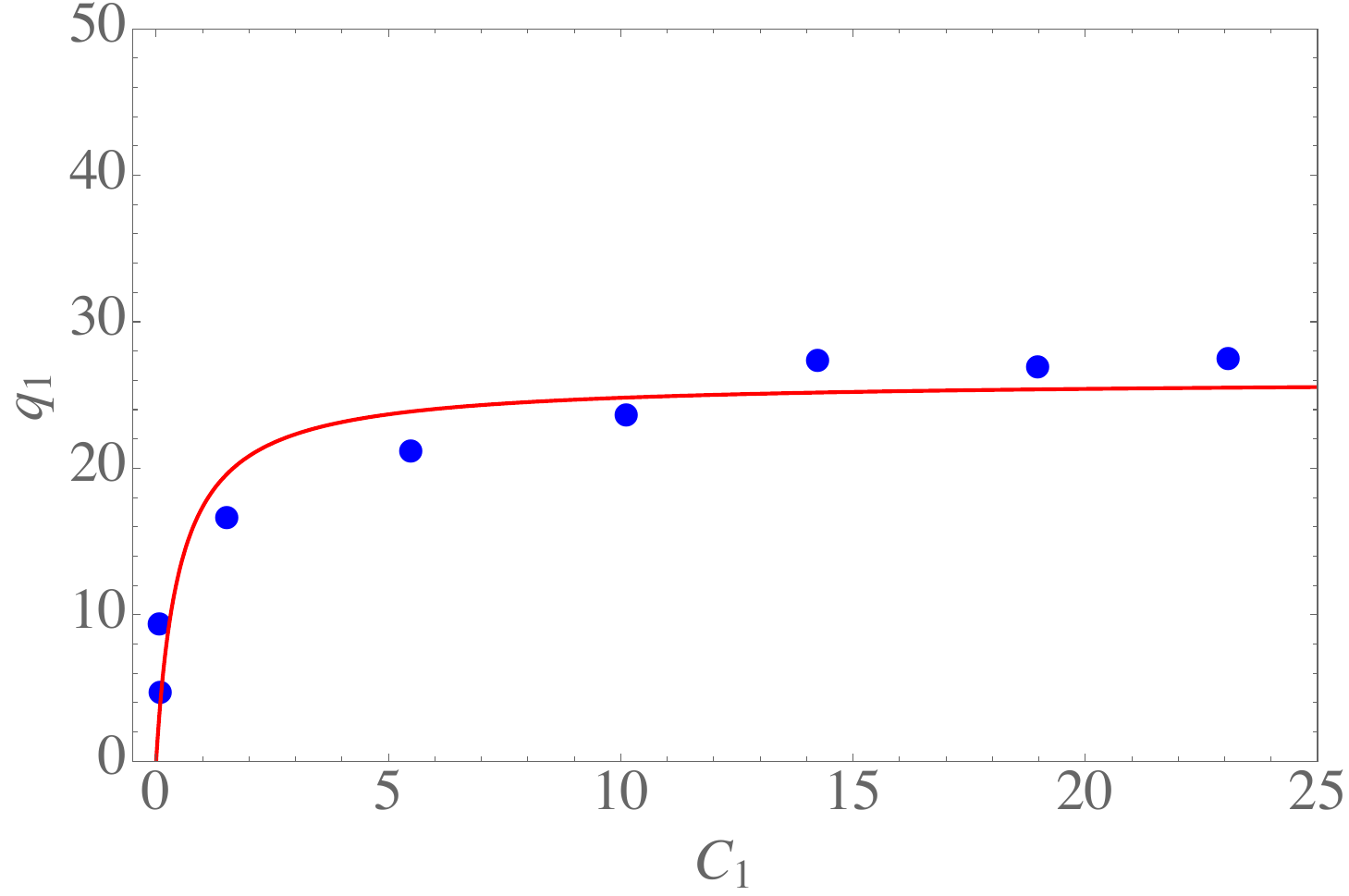}
    \caption{\label{fig:fig3}  Experimental and predicted stable equilibirum isotherms of the dimensionless adsorption capacity versus $C_1$ for As (V) onto magnetic graphene oxide. The blue dots represent the experimental data reported in~\cite{yuEnhancedRemovalPerformance2015}, and the red curve the stable equilibrium solution predicted by the SEAQT equation of motion. }
\end{figure}

For the sake of comparison, a regression fit of the predictions made with the Langmuir model equivalent, i.e., Eq.~(\ref{eq:ads}), of the SEAQT prediction are compared with the experimental data given in~\cite{yuEnhancedRemovalPerformance2015} for the adsorption of As (V) onto magnetized graphene oxide. The results are presented in Fig.~\ref{fig:fig3}. The best fit parameters for the regression analysis correspond to a $\xi_n^{ad}$ of 1.99 and an $M$ of approximately 26.05 with a correlation coefficient $R^2 \approx 0.9829$. The results show good agreement with the experimental values. This analysis is intended only to demonstrate the compatibility of the SEAQT framework with conventional equilibrium adsorption models. However, it is important to emphasize that the SEAQT framework unlike the Langmuir model makes no $a \; priori$ assumption of a kinetic model. Instead, the adsorption isotherms are derived from the energy eigenstructure of the material under consideration using the SEAQT equation of motion. In a companion paper, a detailed derivation of the energy eigenstructure for an adsorption system is derived from a non-Markovian Monte Carlo simulation, which is then used by the SEAQT equation of motion to predict the non-equilibrium and equilibrium adsorption process of As~\cite{saldana-roblesSteepestentropyascentFrameworkPredicting2024}.

As seen in Fig.~\ref{fig:fig3}, as the Langmuir parameter, $C_1$, becomes larger, the dimensionless adsorption increases, as expected. From the SEAQT point of view, this implies that the adsorption could be improved as the absolute activity (or chemical potential) of the adsorbate is increased. For larger concentrations in the limit of  $C_1 \gg b$, the asymptotic limit gives
\begin{equation}
\lim_{C_1 \rightarrow \infty} \frac{q_1}{b}= 1
\end{equation}
i.e., the  adsorption capacity, $q_1$ becomes $\propto M$. Thus, from the SEAQT point of view, the greater the number of particles available for adsorption is, the higher the maximum adsorption capacity. For small concentrations, the fugacity is small and as a consequence, $\xi_n^{ad} \exp( -\beta_R e_1) \ll 1$ so that the Langmuir equation reduces to the expected Henry form of
\begin{equation}
    \frac{\langle n \rangle ^{eq}_{ad}}{M} = g_1\, \xi_n^{ad}\, \exp \left( -\beta_R \, e_1 \right)
\end{equation}
which indicates a linear relationship between the fugacity and the particles adsorbed at equilibrium. Another way to see  that the SEAQT model is compatible with the Henry adsorption model is by considering the equilibrium solution where $\langle \Phi n \rangle_{ad} = \langle f \rangle \langle n \rangle_{ad} - \tilde \mu_n \left(\langle n \rangle_{ad}\right)^2$ , which when $\tilde \mu_n \ll 1$, leads to the Henry equation.

\subsection{Freundlich equation}
The Freundlich model is an empirical model whose theoretical derivation is not well established. This model can be written as
\begin{equation}
    \langle n \rangle^{eq}_{ad} = \alpha \langle n \rangle_0^\delta \left( 1- y \right)^\delta
\end{equation}
where the removal fraction $y = \langle n \rangle^{eq}_{ad} / \langle n \rangle_0$  is defined as the ratio of the average adsorbed particle number at stable equilibrium, $\langle n \rangle^{eq}_{ad}$, to the initial average particle number,  $\langle n \rangle_0$ . The average adsorbed particle number at stable equilibrium is calculated from the grand canonical partition function, and the coefficients $\alpha$ and $\delta$ are related to the Freundlich adsorption capacity and intensity. In the notation of the SEAQT model, equilibrium requires that
\begin{equation} \label{eq:fl}
    \langle \Phi n \rangle_{ad} =  \alpha \langle n \rangle_0^\delta \left( \langle f \rangle (1-y)^{\delta} - \mu (1-y)^{2 \delta} \right)
\end{equation}
where the right-hand side can be shown to be 
\begin{equation}
 \langle \Phi n\rangle_{ad} = -\frac{\ln \mathcal{Z}^{ad}}{\beta_R} \frac{\partial \ln \mathcal{Z}^{ad}}{\partial \tilde \mu_n}
\end{equation}
This is the expected value at stable equilibrium from factoring the adsorbed grand canonical partition function, $\mathcal{Z}^{ad}$, and the average adsorbed particle number. For high removal fractions (($y \approx 1$), Eq.~(\ref{eq:fl}) can be further reduced to
\begin{equation}
     \frac{\ln \mathcal{Z}^{ad}}{\ln Z} \left(\frac{\partial \ln \mathcal{Z}^{ad}}{\partial \tilde \mu_n} \right) = \alpha \langle n \rangle_0^{\delta}\, (1-y)^{\delta}
\end{equation}
Thus, the Freundlich isotherm is compatible with the SEAQT framework for $\mathcal{\mu} \ll 1$ (i.e., $\ln \mathcal{Z}^{ad} \sim \ln  Z$ where $Z$ is the canonical partition function) and high removal fractions.  Furthermore, the SEAQT model leads to an identification of the empirical constants of the Freundlich model with the thermodynamic information of the system at stable equilibrium such as the canonical partition function.

\subsection{Multicomponent Langmuir}
Now, consider the equations of motion in the SEAQT framework with two components $n$ and $m$. The grand potential in this case is
\begin{equation}
    \langle \Phi \rangle = \langle e \rangle - \beta^{-1}_R \langle s \rangle - \mu_{R_n} \langle n \rangle - \mu_{R_m} \langle m \rangle \,,
\end{equation}
where again $\mu_{R_n} = \gamma_n/ \beta_R$ and $\mu_{R_m} = \gamma_m / \beta_R$ are the usual chemical potentials for species $n$ and $m$. In the same manner as was done for the single component Langmuir model,  the grand canonical partition function with only the adsorbed species part and the first three energy levels can be written as 
\begin{align}
    \mathcal{Z}^{ad} &= g_0 \exp \left( -\beta_R e_0 + \gamma_n n_0^{ad} + \gamma_m m_0^{ad} \right) \\
    &+ g_1 \exp \left( -\beta_R e_1 + \gamma_n n_1^{ad} + \gamma_m m_1^{ad} \right) \\
    &+ g_2 \exp \left( -\beta_R e_2 + \gamma_n n_2^{ad} + \gamma_m m_2^{ad} \right)     
\end{align}
where the degenerangies, $g_i$,  correspond to the energy levels, $e_i$,  and $n_i^{ad}$ and $m_i^{ad}$ are the adsorbed occupation particle numbers at the energy levels. 

For simplicity, it is assumed that the energy at which the adsorption takes place for each specie is at different levels, which is plausible for a non-competitive adsorption process. In such a situation, the average adsorption for the $n$ species can be associated with energy level 1, and that for the $m$ species with energy level 2. In this case, the grand canonical partition function can be written as
\begin{eqnarray}
    \mathcal{Z}^{ad} &=& 1 + g_1 \exp \left( -\beta_R\, e_1 + \gamma_n \, n_1^{ad} \right) \nonumber\\ 
    &\;& \; + \; g_2 \exp \left( -\beta_R\, e_2 + \gamma_m \, m_2^{ad} \right) 
\end{eqnarray}
where again the lowest energy level can be set to zero so that the dimensionless chemical potential is zero. Using the absolute activity for each species, $\xi_n^{ad}= \exp ( \gamma_n n_1^{ad})$ and $\xi_m^{ad}= \exp ( \gamma_m m_1^{ad})$, the partition function is rewritten as 
\begin{equation}
    \mathcal{Z}^{ad} = 1 + g_1 \exp \left( -\beta_R \, e_1 \right) \xi_n^{ad}+ g_2 \exp \left( -\beta_R \, e_2 \right) \xi_m^{ad}
\end{equation}
The average number of $n$ species particles adsorbed for the $M$ collections of non-interacting particles can then be written as
\begin{equation}
 \frac{\langle n \rangle^{eq}_{ad} }{M} = \frac{ g_1 e^{\left( -\beta_R\, e_1 \right) \xi_n^{ad}}}{1 + g_1 e^{\left( -\beta_R\, e_1 \right) \xi_n^{ad}}+ g_2 e^{\left( -\beta_R\, e_2 \right) \xi_m^{ad}}}   
\end{equation}
which is compatible with the generalized Langmuir isotherm model~\cite{amruthaMulticomponentAdsorptionIsotherms2023} with $C_i= g_i \exp \left( -\beta_R e_i\right)$ and $b_i = \xi_i^{ad}$. 

If instead a competitive adsorption process occurs, which energetically can be understood as the $m$ and $n$ species being adsorbed at the same energy level, the grand canonical partition function takes the form
\begin{eqnarray}
  \mathcal{Z}^{ad} &=&  1 + g_1 \exp \left( -\beta_R \,e_1 \right) \xi_n^{ad}\, \xi_m^{ad} \nonumber \\
  &\;& \; +\; g_1 \exp \left( -\beta_R\, e_2 \right) \xi_n^{ad}\, \xi_m^{ad}     
\end{eqnarray}
and the average number of $n$ species particles adsorbed becomes 
\begin{equation}
 \frac{\langle n \rangle^{eq}_{ad} }{M} = \frac{ g_1 e^{\left( -\beta_R\, e_1 \right) \xi_n^{ad} \, \xi_m^{ad}}}{1 + g_1 e^{\left( -\beta_R\, e_1 \right) \xi_n^{ad}\, \xi_m^{ad}}+ g_1 e^{\left( -\beta_R\, e_2 \right) \xi_n^{ad}\, \xi_m^{ad}}}   
\end{equation}
which is compatible with the competitive Langmuir model where again $C_i= g_i \exp \left( -\beta_R e_i\right)$ and $b_i = \xi_i^{ad}$. 

\section{Conclusion}
\label{sec:con}

This work shows how the mathematics of the SEAQT framework can be used to derive an equation of motion for predicting both the non-equilibrium and stable equilibrium adsorption behavior of systems with one or more species. It also shows how the stable equilibrium solution predicted by this framework is consistent with the standard adsorption isotherm models of Langmuir and Freundlich. The practical application of this framework to the absorption process of As on graphene oxide at both non-equilibrium and stable equilibrium is provided in a companion paper \cite{saldana-roblesSteepestentropyascentFrameworkPredicting2024}.

To address comprehensively both the adsorbate and sorbent behaviors, the SEAQT equation of motion incorporates a number operator associated with each adsorbed and unadsorbed species in the system. The outcome reveals that, with the inclusion of these operators, the equilibrium solution of the SEAQT framework aligns with the grand-canonical distribution. This alignment offers a quantum-thermodynamic interpretation of, for example, the Langmuir parameters at stable equilibrium, elucidating their connection with the generalized, i.e., non-equilibrium and stable equilibrium, Massieu function and generalized thermodynamic intensive properties $\beta$ and $\gamma_i$. This quantum thermodynamic perspective adds a layer of depth to our understanding of the kinetics of the adsorption phenomenon and its behavior at stable equilibrium.

\appendix
\section{Fluctuations at equilibrium}
\label{sec:fluctuation}
The binary fluctuations, $A_{xy}$, for any two extensive variables $x$ and $y$ can be written in terms of derivatives of the logarithm of the partition function by considering its conjugate variables $\tilde x$ and $\tilde y$. Namely, for the extensive variable $\langle e \rangle$ (energy), the conjugate intensive variable is $\beta$ (reciprocal temperature). Thus, in the grand canonical ensemble, the fluctuation of energy is simply given by
\begin{equation} \label{eq:Aee}
A_{ee} = \frac{\partial^2 }{\partial \beta^2} \ln \mathcal{Z}
\end{equation}
where at equilibrium $\ln \mathcal{Z} = -\beta \langle \Phi \rangle_{\text{eq}}$. In the same manner, the binary fluctuation of particle number is given as
\begin{equation} \label{eq:Ann}
    A_{nn} = \frac{\partial^2 }{\partial \gamma_n} \ln \mathcal{Z}
\end{equation}
where the conjugate intensive variable of the dimensionless chemical potential, $\gamma_n$, equals $\beta \mu_n$ and $\mu_n$ is the dimensional chemical potential. This structure follows from the Boltzmann factor that appears in the partition function, which can be written as the product of two exponentials that depend upon the product of the intensive properties $\beta$ and $\tilde \mu_n$. Furthermore, because of this structure, the fluctuation of energy and particle number is simply
\begin{equation} \label{eq:Aen}
    A_{en} = \frac{\partial^2}{\partial \beta \,\partial \gamma_n} \ln \mathcal{Z} 
\end{equation}
which allows one to conclude that the fluctuations $A_{en} = A_{ne}$ are symmetric. 

The entropy fluctuations can be written as
\begin{equation} \label{eq:Ass}
    A_{ss} = \beta^2 \frac{\partial^2 }{\partial \beta^2} \ln \mathcal{Z}
\end{equation}
while the energy-entropy fluctuation is given by
\begin{equation} \label{eq:Aes}
A_{es} = \beta \frac{\partial^2 }{\partial \beta^2} \ln \mathcal{Z}
\end{equation}
Thus, the ratio of Eq.~(\ref{eq:Aes}) to Eq.~(\ref{eq:Ass}) results at stable equilibrium in the temperature~\cite{berettaNonlinearQuantumEvolution2009}. 

This structure can be generalized to fluctuations among three extensive variables as a determinant of the kind 
\begin{equation} \label{eq:Aenn}
    \det \text{Hess} \left( \beta \langle \Phi \rangle_{\text{eq}} \right) = \det \left|
\begin{matrix}
A_{ee} & A_{en}  \\
A_{en} & A_{nn}
\end{matrix}
    \right|
\end{equation}
By directly substituting Eqs. (\ref{eq:Aen}), (\ref{eq:Ann}), and (\ref{eq:Aee}) into Eq.~(\ref{eq:Aenn}), the determinant of the Hessian of the logarithm of the partition function can be expressed as a function of the intensive properties $\beta$ and $\gamma_n$ since $\ln \mathcal{Z} = \ln \mathcal{Z}(\beta, \gamma_n)$. From the construction of the SEAQT equation of motion, this determinant is identified as a Gram determinant, which is inherently positive definite. Therefore, it follows that at stable equilibrium, the determinant of the Hessian is positive definite. This implies that the system reaches a maximum of the logarithm of the grand canonical partition function. Given that the logarithm of this partition function is related to the free energy, i.e., $\beta \langle f \rangle = -\ln Z$ where $Z$ is the canonical partition function, then the generalized fluctuation theorem implies the minimization of the free energy as expected.

\bibliography{references_z}

\end{document}